\documentclass[preprintnumbers,showpacs,amsmath,amssymb,floatfix,prd,onecolumn,superscriptaddress,nofootinbib]{revtex4}
\usepackage{amssymb}
\usepackage{bbm}
\usepackage{graphicx}
\usepackage{epsfig}
\usepackage{bm}
\usepackage{amsfonts}
\usepackage{epstopdf}
\usepackage{multirow}

\begin{document}

\title{Latest Observational Constraints to the Ghost Dark Energy Model  \\by Using Markov Chain Monte Carlo Approach }

\author{Chao-Jun Feng}
\email{fengcj@shnu.edu.cn} \affiliation{Shanghai United Center for Astrophysics (SUCA), \\ Shanghai Normal University,
    100 Guilin Road, Shanghai 200234, P.R.China}

\author{Xin-Zhou Li}
\email{kychz@shnu.edu.cn} \affiliation{Shanghai United Center for Astrophysics (SUCA),  \\ Shanghai Normal University,
    100 Guilin Road, Shanghai 200234, P.R.China}

\author{Xian-Yong Shen}
\email{1000304237@smail.shnu.edu.cn } \affiliation{Shanghai United Center for Astrophysics (SUCA), \\ Shanghai Normal University,
    100 Guilin Road, Shanghai 200234, P.R.China}

\begin{abstract}
Recently,  the vacuum energy of the QCD ghost in a time-dependent background is proposed as a kind of dark energy candidate  to explain the acceleration of the universe. In this model, the energy density of the dark energy   is proportional to the Hubble parameter $H$, which is  the Hawking temperature on the Hubble horizon of the Friedmann-Robertson-Walker (FRW) universe. In this paper, we perform a constraint on the ghost dark energy model with and without bulk viscosity, by using the Markov Chain Monte Carlo (MCMC) method and the combined latest observational data from the type Ia supernova compilations including Union2.1(580) and Union2(557), cosmic microwave background, baryon acoustic oscillation, and the observational Hubble parameter data.
\end{abstract}


\maketitle


\section{Introduction}\label{sec:intro}

Since the time people find that the universe is accelerated expanding,  lots of theoretical  models were proposed to explain this phenomenology. In some of them, people proposed a new kind of dark energy component with negative pressure  in our university, which will drive the acceleration, and the simplest dark energy model is the cosmological constant, but it suffers fine-tuning and coincidence problems. While, in other models, people try to modify the Einstein gravity at large scale in the universe, e.g. $f(R)$, DGP, etc. models, then the universe can be accelerated without introducing dark energy.

Recently, a very interesting dark energy model called Veneziano ghost dark energy (GDE) has been proposed \cite{Urban}, and in this model, one can obtain a cosmological constant of just the right magnitude to give the observed expansion from the contribution of the ghost fields, which are supposed to be present in the low-energy effective theory of QCD without introducing any new degrees of freedom.  The ghosts are needed to solve the $U(1)$ problem, but they are completely decoupled from the physical sector \cite{Kawarabayashi:1980dp}. The ghosts make no contribution in the flat Minkowski space, but make a small energy density contribution to the vacuum energy due to the off-set of the cancelation of their contribution in curved space or time-dependent background. For, example, in the Rindler space, the contribution of high frequency modes is suppressed by the factor $e^{-2\pi k/a}$ and the main contribution comes from $k\sim a$, where $a$ is the temperature on the horizon seen by the Rindler observer \cite{Ohta}. In the cosmological context,  one can choice $a\sim H$, which corresponds to the temperature on the Hubble horizon. Then, in the context of strongly interacting confining QCD with topological nontrivial sector, this effect occurs only in the time direction and their wave function in other space directions is expected to have the size of QCD energy scale. Thus, this ghost gives the vacuum energy density proportional to $\Lambda_{QCD}^{3} H_{0}$. With  $H_{0}\sim 10^{-33}$eV, it gives the right order of observed magnitude $\sim (3\times10^{-3}$eV)$^{4}$ of the energy density. The recent progress on the GDE model, see Ref.~\cite{Cai:2010uf, Feng:2011ev, Ebrahimi:2011js}, in which the author also consider other possibilities of the energy density form of the GDE model with or without interactions and also its thermodynamical behaviors.

Dissipative processes in the universe including bulk viscosity, shear viscosity and heat transport have been
conscientiously studied\cite{barrow}, and these dissipative effects are supposed to play a very significant role in the astrophysics \cite{astrop} and nuclear physics \cite{nuclear}. Therefore, it is also important to study these effects in the GDE model, and in this paper, we will consider the GDE model with and without bulk viscosity and  constrain the relevant parameters in these models by using the latest data, and for works on viscous dark energy models, see ref.\cite{vde}. The general theory of dissipation in relativistic imperfect fluid was put on a
firm foundation by Eckart\cite{eckart}, and, in a somewhat different formulation, by Landau and Lifshitz\cite{landau}.
This is only the first order deviation from equilibrium and may has a causality problem, the full causal theory was
developed by Isreal and Stewart\cite{israel}, and has also been studied in the evolution of the early
universe\cite{harko}. However, the character of the evolution equation is very complicated in the full causal theory.
Fortunately, once the phenomena are quasi-stationary, namely slowly varying on space and time scale characterized by
the mean free path and the mean collision time of the fluid particles, the conventional theory is still valid. In the
case of isotropic and homogeneous universe, the dissipative process can be modeled as a bulk viscosity $\zeta$ within a
thermodynamical approach, while the shear viscosity $\eta$ can be neglected, which is consistent with the usual
practice\cite{brevik}.

The bulk viscosity introduces dissipation by only redefining the effective pressure, $p_{eff}$, according to
$p_{eff}=p-3\xi H$ where $\xi$ is the bulk viscosity coefficient and $H$ is the Hubble parameter. The case $\xi =\tau H$, implying the bulk viscosity is proportional to the fluid's
velocity vector, is physically natural, and has been considered earlier in  a astrophysical context, see the review
article of Gr{\o}n\cite{gron}.

In this paper, we will perform an observational constraint on the parameters of two kinds of GDE models by using  the Markov Chain Monte Carlo (MCMC) method. One of them is the original model without bulk viscosity (GDE), and the other is the one with  bulk viscosity (VGDE). The observational data set we used include the latest update of type Ia supernova data "Union2.1" with 580 SNe \cite{Suzuki:2011hu}, and for comparison we also used the "Union2" data with 557 SNe \cite{Amanullah:2010vv}. Also, we used  cosmic microwave background (CMB) from WMAP7 \cite{Komatsu:2010fb}, baryon acoustic oscillations (BAO) from SDSS DR7 \cite{Percival:2009xn} and the observational Hubble parameter data (OHD) from \cite{Stern:2009ep,Riess:2009pu,Gaztanaga:2008xz,Simon:2004tf} for combination constraint. We also added a new parameter $\tau$ in the VGDE model and modified the public available \textbf{CosmoMC} package \cite{Lewis:2002ah} to satisfy both kinds of these models. During the calculation, we run $32$ independent chains for each case, and we make sure the convergence of the chains by typically getting $\mathcal{R}-1$ to be less than $0.01$, see Ref.~\cite{Lewis:2002ah} and the readme file on the website of \textbf{CosmoMC}.

This paper is organized as follows: In Section \ref{sec:background}, we will derive the background dynamical equations for both the VGDE and GDE model. In Section \ref{sec:auto}, we will  study the future behaviors of the VGDE model by using the dynamical analysis. In Section \ref{sec:data}, we will describe the method and data, and in Section \ref{sec:result}, we summarize the fitting results.  In the last section, we will give some discussions and conclusions.

\section{Background solution}\label{sec:background}

The energy density of the ghost dark energy is proposed as \cite{Urban}
\begin{equation}
	\rho_{de} = \alpha H \,,
\end{equation}
and the Friedmann equation for a flat universe  is given by
\begin{equation}\label{fri}
	H^{2} = \frac{\rho}{3} \,, \quad \rho = \rho_{de} + \rho_{m} + \rho_{r} \,,
\end{equation}
where we have set $8\pi G=1$.
By solving the above equation, we obtain
\begin{equation}\label{sol}
	\rho_{de} =  \frac{\alpha^{2}}{6} \left(1+\sqrt{1+\frac{12\tilde\rho_{m}}{\alpha^{2}}}\right) \,,
\end{equation}
where $\tilde\rho_m = \rho_m + \rho_r$.
The continuity equations for the matter $\rho_m$ and total energy density $\rho$ are given by
\begin{eqnarray}
	\dot{\tilde{\rho} }_{m}&=& -3H\left(\tilde\rho_{m} + \frac{\rho_r}{3} - 3\xi_m H \right) \,, \label{con1} \\
	\dot \rho &=& -3H(\rho + p - 3\xi H) \,,       \label{con2}
\end{eqnarray}
where we have neglected the pressure of matter $p_m\approx 0$. Here $\xi$ and $\xi_m$ are the bulk viscosity for the total energy density and matter respectively, so the evolution behavior of radiation is still as usual, i.e. $\rho_r\sim a^{-4}$ . The pressure $p$ is defined as the effective pressure of total fluid without bulk viscosity, and the first law of thermodynamics in an adiabatic expanding universe gives
\begin{equation}
  p= \left(\tilde\rho_m + \frac{\rho_r}{3}\right)\frac{\partial \rho}{\partial\tilde\rho_m} - \rho \,.
\end{equation}
Thus, one can get $\xi = \frac{\partial\rho}{\partial\tilde\rho_m}\xi_m$ from Eqs.~(\ref{con1}) and (\ref{con2}). In the following, we will choose $\xi = \tau H$, in which the cosmological dynamics can be analytically solvable [19] and $\tau$ is a constant. By using the solution (\ref{sol}),
 the Friedmann equation can be rewritten as
\begin{equation}\label{fri2}
    E^2 \equiv \frac{H^2}{H^2_0} = (\Omega_{m0} +\Omega_{r0}) f(y)F \,,
\end{equation}
where $F = \tilde\rho_{m}/\tilde\rho_{m0}$, and we have defined  $\Omega_{i0}=\rho_{i0}/\rho_{c0}, (i = m, r)$, in which the subscript zero denotes present values. Here, $f = \rho/\tilde\rho_{m}$ and  $y = \ln(\tilde\rho_{m}/\alpha^{2})$, and from Eq.~(\ref{sol}), we get
\begin{equation}
	f(y)= 1+ \frac{1+ \sqrt{1+12e^{y} }}{6e^{y}} \,,
\end{equation}
and then we have the constraint from the Friedmann equation
\begin{equation}\label{cons}
	f(y_0) = \frac{1}{\Omega_{m0} +\Omega_{r0}} \,, \quad \text{or} \quad
e^{y_{0}} = \frac{\Omega_{m0} +\Omega_{r0}}{3(1-\Omega_{m0} -\Omega_{r0})^{2}} \,,
\end{equation}
where $y_{0} = \ln(\tilde\rho_{m0}/\alpha^{2})$.
The conservation law (\ref{con2}) becomes
\begin{equation}
  f' +3f(1-f)\left(\frac{1}{1+f} - \frac{\tau }{2} \right) + \frac{\Omega_{r0}}{(1-\Omega_{m0} -\Omega_{r0})^{2}} \frac{(1-f)^3}{1+f} e^{-4x} = 0 \,,
\end{equation}
where the prime denotes the derivative with respect to $x\equiv \ln a = -\ln(1+z)$, and $z$ is the redshift. Of course, the above equation is not easy to solve analytically, but once one gets the solution of $f$ by using the numerical method for example, then, the Hubble parameter (\ref{fri2}) is given by
\begin{equation}
  E = \left(1-\Omega_{m0} -\Omega_{r0}\right)\left[ 1+\frac{1}{f(x)-1} \right] \,.
\end{equation}

However, in the later time ($x\sim 0$), the contribution of radiation could be neglected, and one can get the following solution
\begin{eqnarray}
\nonumber
	\rho_{m} &=& \rho_{m0}(1+z)^{3(1-\tau)}  \left[ \frac{12(1-\tau)^{2} +\tau(\tau-2) e^{-y_0}F^{-1}}{12(1-\tau)^{2} +\tau(\tau-2) e^{-y_{0}}}\right]^{\frac{1}{\tau -2}} \\
\nonumber
	&\cdot& \exp\left[ \frac{2(1-\tau)}{(\tau-2)} \bigg(\arctan \sqrt{1+12e^{y_0}F}-\arctan\sqrt{1+12e^{y_{0}}} \bigg)\right]\\
	&\cdot& \exp\left[ \frac{2}{(\tau-2)} \bigg(\arctan\left [(\tau-1)\sqrt{1+12e^{y_0}F}\right ]-\arctan\left[(\tau-1)\sqrt{1+12e^{y_{0}}}\right] \bigg)\right] \,,
\end{eqnarray}
and the constraint (\ref{cons}) reduces to
\begin{equation}\label{cons red}
	e^{y_{0}} = \frac{\Omega_{m0}}{3(1-\Omega_{m0})^{2}} \,.
\end{equation}
Furthermore, if $\tau = 0$, the solution is rather simple
\begin{equation}
	E^{2}|_{\tau=0} =   \Omega_{m0}(1+z)^{3}
	+\frac{1}{2 } \left[ (1-\Omega_{m0})^{2}+  (1-\Omega_{m0})\sqrt{ (1-\Omega_{m0})^{2}+4 \Omega_{m0}(1+z)^{3} } \right] \,.
\end{equation}
Then, the total effective equation of state is given by
\begin{equation}
	w_{eff} = \frac{p}{\rho} = - \frac{e^{-y_{0}}}{6F} \left(1+\frac{1+6e^{y_{0}}F}{ \sqrt{1+12e^{y_{0}}F}}\right) \,.
\end{equation}
In the case of $\tau=0$,
\begin{equation}
	w_{eff}|_{\tau=0} =- \frac{(1-\Omega_{m0})^{2}}{2 \Omega_{m0}(1+z)^{3}}
	\left[1+\frac{(1-\Omega_{m0})^{2}+2 \Omega_{m0}(1+z)^{3}}
	{ (1-\Omega_{m0})\sqrt{(1-\Omega_{m0})^{2}+4 \Omega_{m0}(1+z)^{3}}}\right] \,.
\end{equation}
In the following, we will neglect the contribution of radiation, namely $\Omega_{r0}\ll \Omega_{m0}$ until doing the parameter fitting from observations.

\section{Autonomous system}\label{sec:auto}

A general study of the phase space system of quintessence and phantom in FRW universe has been given in Ref.[20]. For our model with assumption $\xi = \tau H^{2\beta+3} \sim \rho^{\beta + \frac{3}{2}}$ with $\tau >0 $ and $\beta\neq-1$, the dynamical system could be described by
\begin{eqnarray}
	\dot H &=& - \frac{1}{2} \left( \rho_{m} \frac{d\rho}{d\rho_{m}} - 3\xi H \right) \,, \\
	\dot \rho_{m} &=&  -3H\left(\rho_{m} - 3\xi \frac{d\rho_{m}}{d\rho} H\right) \,.
\end{eqnarray}
To analyze the dynamical system, we rewrite the equations with the
following dimensionless variables:
\begin{equation}
	u= \frac{\rho_{m}}{3H^{2}} \,, \quad v = \frac{1}{\beta +1}\ln \left( \frac{\xi}{H} \right) \,,
\end{equation}
then,  The dynamical system can be reduced to
\begin{eqnarray}
	u' &=&  3g(u,v)^{-1}\bigg[ u g(u,v)-1\bigg]\bigg[ ug(u,v)-e^{(\beta+1)v}\bigg] \,, \label{d1} \\
	v' &=& -3 \bigg[u g(u,v) - e^{(\beta+1)v} \bigg]\,, \label{d2}
\end{eqnarray}
where
\begin{equation}
	g(u,v) = 1+ \bigg(1+\tilde \tau u e^{v}\bigg)^{-\frac{1}{2}} \,, \quad \tilde \tau \equiv 36\alpha^{-2} \tau^{-\frac{1}{\beta+1}} \,,
\end{equation}
thus, we have $1<g<2$. And we also have a constraint from Friedmann equation
\begin{equation}
	1 =  \frac{ug}{2-g} \,.
\end{equation}

In the following, we will study the dynamical system for the variables $\bold\Omega\equiv(u, v)$, determined by the Eq.~(\ref{d1}) and (\ref{d2}). As we known, for a autonomous system $\bold \Omega = f(\bold\Omega)$, where $f$ can be extracted from the above equations (\ref{d1}) and (\ref{d2}), the critical points $\bold\Omega^{*}$ of the system are given by the condition $f(\bold\Omega^{*}) = 0$. Thus, all the critical points $(u_{c},v_{c})$  in our model satisfy $u_{c} g(u_{c},v_{c})=e^{(\beta+1)v_{c}}$.  To investigate the stability of the critical points, we will expand the autonomous system around the critical points $\bold\Omega = \bold\Omega^{*} + \delta\bold\Omega$ with the perturbation $\delta\bold\Omega$, and thus we get the perturbation equations at the critical point as $\delta\bold\Omega' = \bold A \delta\bold\Omega$, in which $\bold A$ is the Jacobian matrix evaluated at the corresponding critical point,
\begin{equation}       
\bold A= \frac{3(2-g_{c})}{2}
\left(
\begin{array}{ccc}
1-g_{c}^{2} && g_{c}^{-1}(g_{c}-1)\bigg[ 2\beta+2+ g_{c}(g_{c}-1)\bigg] \\
 \\
 -g_{c} (1+g_{c}) && 2\beta+2+ g_{c}(g_{c}-1)
\end{array}
\right)\,.
\end{equation}
where $g_{c} = g(u_{c},v_{c})$, and here we have $\det \bold{A} = 0$. Thus, the corresponding eigenvalues are
\begin{equation}
	\lambda_{1} = \frac{3(2-g_{c})}{2}\bigg[ 2\beta +3-g_{c}\bigg] \,, \quad
	\lambda_{2} = 0\,,
\end{equation}
Thus, when $\beta < -1$, all the critical points are stable, while $\beta > -1/2$, all of them are unstable, otherwise, only parts of them are stable.

\section{Statistical analysis with the observation data}\label{sec:data}

In this section, we will use the Markov Chain Monte Carlo (MCMC) method to make a fitting on the cosmological parameters in the GDE and VGDE models. We have modified the public available \textbf{CosmoMC} package \cite{Lewis:2002ah} to satisfy this dark energy model, in which we have added a new parameter $\tau$ with a prior $\tau \in [-2.0,2.0]$ in the VGDE model. Thus, in GDE model, the parameter space is $(\Omega_{b0}h^2, \Omega_{m0}h^2, H_0)$, while in the VGDE model, the parameter space is $(\Omega_{b0}h^2, \Omega_{m0}h^2, H_0, \tau)$. The prior for the physical baryon density, dark matter energy density and Hubble constant are $\Omega_{b0}h^2 \in [0.005 0.1], \Omega_{m0}h^2 \in [0.01 0.99], \text{and}~h\in[0.4, 1.0]$. We also impose a weak Gaussian prior on the physical baryon density $\Omega_{b0}h^2 = 0.022\pm 0.002$ from the Big Bang nucleosynthesis \cite{Burles:2000zk}. In this paper, we will use latest data from supernovae of type Ia (SN) \cite{Suzuki:2011hu, Amanullah:2010vv} , current cosmic microwave background (CMB) \cite{Komatsu:2010fb}, baryon acoustic oscillations (BAO) \cite{Percival:2009xn} and the observational Hubble parameter data (OHD) \cite{Stern:2009ep,Riess:2009pu,Gaztanaga:2008xz,Simon:2004tf} to performing the fitting procedure. All the calculations are performed on the Astrophysical Beowulf Cluster (\textbf{AstroBC}), which is a parallel computing network system designed, set up and tested by ourselves, and it is much fast and stable than a personal computer.

In order to determine the best value of parameters (with $1\sigma$  error at least ) in the dark energy model, we
will use the maximum likelihood method and we will take the total likelihood function $\mathcal{L} = e^{-\chi^2/2}$ as the products of these separate likelihood functions of each data set, thus we get
\begin{equation}\label{tot chi2}
    \chi^2 = \chi^2_{SN} +\chi^2_{CMB} + \chi^2_{BAO} + \chi^2_{H}\,.
\end{equation}
Then, one can get the best fitting values of parameters by  minimizing $\chi^2$.

\subsection{Type Ia Supernovae Data and Constraint}
In general, the expansion history of the universe $H(z)$ or $E(z)$ can be given by a specific cosmological model or by
assuming an arbitrary ansatz, which may be not physically motivated but just designed to give a good fit to the data
for the luminosity distance $d_L$ or the 'Hubble-constant free' luminosity distance $D_L$ defined by
\begin{equation}\label{lu dis}
    D_L = \frac{H_0 d_L}{c} \,,
\end{equation}
where the light speed $c$ is recovered to show that $D_L$ is dimensionless. In the following, we will take the first
strategy that assuming the Hubble parameter $H(z; a_1, \cdots, a_n)$ with some parameters ($a_1,\cdots, a_n$) predicted by
the viscous ghost dark energy model could be used to describe the universe, and then we obtain the predicted value of $D^{th}_L$ by
\begin{equation}\label{Dlth}
    D^{th}_L = (1+z)\int^z_0  \frac{dz'}{E(z';a_1, \cdots, a_n)}\,,
\end{equation}
for a flat universe. On the other hand, the apparent magnitude of the supernova is related to the corresponding luminosity distance by
\begin{equation}\label{app mag}
   \mu(z) = m(z) - M = 5\log_{10}\left[\frac{d_L(z)}{\text{Mpc}}\right] + 25 \,,
\end{equation}
where $\mu(z)$ is the distance modulus and  $M$ is the absolute magnitude which is assumed to be constant for standard
candles like Type Ia supernovae. One can also rewrite the distance modulus in terms of $D_L$ as
\begin{equation}\label{app mag2}
    \mu(z) = 5\log_{10} D_L(z) + \mu_0 \,,
\end{equation}
where
\begin{equation}\label{zero off}
   \mu_0 = 5\log_{10} \left(\frac{cH_0^{-1}}{\text{Mpc}}\right) + 25 = -5\log_{10} h + 42.38 \,,
\end{equation}
is the zero point offset, which is an additional model independent parameter. Thus, we obtain the predicted value of
$\mu^{th}$ by using the value of $D_L^{th}$ and the corresponding data sets we used are the "Uinon2.1" SN Ia compilation including 580 SNe \cite{Suzuki:2011hu}, released recently by the Supernova Cosmology Project (SCP) collaboration. It is a  an update of the "Union2" compilation, which contains 557 SNe \cite{Amanullah:2010vv}. For comparison, we will use both of the two data sets, and the $\chi^2$ for the SN Ia is given by
\begin{equation}\label{chi2}
    \chi^2_{SN}(a_1,\cdots,a_n)
    = \sum^{N}_{i,j=1}\bigg[\mu^{obs}(z_i) - \mu^{th}(z_i)\bigg] \bigg(C_{SN}^{-1}\bigg)_{ij} \bigg[\mu^{obs}(z_i) - \mu^{th}(z_i)\bigg]\,,
\end{equation}
where $N = 580 \, \text{and} \,557$ for Union2.1 and Union2 respectively. Here $C_{SN}$ is the covariance matrix with systematic errors.

\subsection{Cosmic Microwave Background  Data and Constraint}
The temperature power spectrum of CMB is sensitive to the physics at the decoupling epoch and the physics between now and the decoupling epoch. The former primarily affects the ratios of the peak heights and the Silk damping, i.e. the amplitude of acoustic peaks, while the latter changes the locations of peaks via the angular diameter distance out to the decoupling epoch. We can quantify this by the "acoustic scale" $l_A$ and the "shift parameter" $R$, which are defined as \cite{Bond:1997wr}
\begin{equation}
  l_A = \frac{\pi r(z_*)}{r_s(z_*)} \,, \quad
  R = \sqrt{ \frac{\Omega_{m0}}{c} }\int^{z_*}_0  \frac{dz'}{E(z')}
\end{equation}
where $z_*$ is the redshift of decoupling. Here $r(z)$ is the comoving distance
\begin{equation}
  r(z) = \frac{c}{H_0} \int^{z}_0  \frac{dz'}{E(z')} \,,
\end{equation}
and $r_s(z_*)$ is the comoving sound horizon at recombination defined by
\begin{equation}
  r_s(z_*) = \int_0^{a(z_*)} \frac{c_s(a)}{a^2 H(a)} da \,,
\end{equation}
where the sound speed $c_s(a)$ is given by
\begin{equation}
  c_s(a) = \bigg[ 3 \left( 1+ \frac{3\Omega_{b0}}{4\Omega_{\gamma 0}} a \right) \bigg]^{-1/2} \,.
\end{equation}
In this paper, we fix $\Omega_{\gamma0} = 2.469\times 10^{-5} h^{-2}$, which is the best fit values given by the seven-year WMAP observations \cite{Komatsu:2010fb}. And the total radiation energy density is the sum of photons and relativistic neutrinos, namely
\begin{equation}
  \Omega_{r0} = \Omega_{\gamma0}(1+0.2271 N_{\text{eff}}) \,,
\end{equation}
where $ N_{\text{eff}}$ is the effective number of neutrino species, and the current standard value is $ N_{\text{eff}}=3.04$ \cite{Mangano:2001iu}. Furthermore, we will use the fitting function of $z_*$ proposed by Hu and Sugiyama \cite{Hu:1995en}:
\begin{equation}
  z_* = 1048 \left[1+0.00124(\Omega_{b0}h^2)^{-0.738}\right]\left[ 1+ g_1(\Omega_{m0}h^2)^{g_2}\right] \,,
\end{equation}
where
\begin{equation}
  g_1 = \frac{0.0783(\Omega_{b0}h^2)^{-0.238}}{1+39.5(\Omega_{b0}h^2)^{0.763}} \,, \quad
  g_2 = \frac{0.560}{1+21.1(\Omega_{b0}h^2)^{1.81} } \,.
\end{equation}
The $\chi^2 $ of the CMB data is constructed as $\chi^2_{CMB} =X^{T}C_{CMB}^{-1}X$, where
\begin{equation}
  X=\left(\begin{array}{l}
    l_{A}-302.09\\
    R-1.725\\
    z_{*}-1091.3\end{array}\right)\,,
\end{equation}
and the inverse covariance matrix is given by \cite{Komatsu:2010fb}
\begin{eqnarray*}
C_{CMB}^{-1} & = & \left(\begin{array}{ccc}
    2.305 & 29.698 & -1.333\\
    29.698 & 6825.270 & -113.180\\
    -1.333 & -113.180 & 3.414\end{array}\right)\, .
\end{eqnarray*}

\subsection{Baryon Acoustic Oscillations Data and Constraint}
We also use the spectroscopic Sloan Digital Sky Survey (SDSS) Data Release 7 (DR7) \cite{Percival:2009xn} to constrain our dark energy model. As we known, for a redshift survey in a thin shell, the position of the BAO approximately constrains $d_z \equiv r_s(z_d)/D_V(z)$, where $r_s(z_d)$ is the comoving sound horizon at the baryon drag epoch, and the definition of $D_V(z)$ is given by \cite{Eisenstein:2005su}
\begin{equation}
  D_V(z) \equiv \bigg[ \left( \int^{z}_0  \frac{dz'}{H(z')} \right)^2\frac{ c z }{H(z)}\bigg]^{1/3} \,.
\end{equation}
The redshift of the drag epoch can be given by the following fitting formula
\begin{equation}
  z_d = \frac{ 1291(\Omega_{m0}h^2)^{0.251} }{ 1+ 0.659(\Omega_{m0}h^2)^{0.828} } \bigg[ 1+ b_1(\Omega_{b0}h^2)^{b_2}\bigg] \,,
\end{equation}
which is proposed by Eisenstein and Hu \cite{Eisenstein:1997ik}. Here
\begin{equation}
  b_1 = 0.313(\Omega_{m0}h^2)^{-0.419} \left[1 + 0.607(\Omega_{m0}h^2)^{0.674}\right] \,, \quad
  b_2 = 0.238(\Omega_{m0}h^2)^{0.223}  \,.
\end{equation}
The $\chi^2 $ of the BAO data is constructed as $\chi^2_{BAO} =Y^{T}C_{BAO}^{-1}Y$, where
\begin{equation}
  Y=\left(\begin{array}{l}
    d_{0.2}-0.1905\\
    d_{0.35}-0.1097\end{array}\right)\,,
\end{equation}
and the corresponding covariance matrix is given by \cite{Percival:2009xn}
\begin{eqnarray*}
C_{BAO}^{-1} & = & \left(\begin{array}{ccc}
    30124 & -17227 \\
    -17227 & 86977\end{array}\right)\,.
\end{eqnarray*}

\subsection{Observational Hubble Parameter Data and Constraint}
Based on different ages of galaxies, the observational Hubble parameter data can be obtained, see Ref.~\cite{Jimenez:2001gg}. It can be used as an independent  constraint on the cosmological parameters. From the Gemini Deep Deep Survey (GDDS) and archival data, Simon et al. obtain $H(z)$ in the range of $0.1\lesssim z \lesssim 1.8 $ \cite{Simon:2004tf}. Stern et al. have used the new data at $0.35\lesssim z \lesssim 1.0 $ from Keck observations, SPICES survey and VVDS survey \cite{Stern:2009ep}. By observing 240 long-period Cepheids \cite{Riess:2009pu}, one can obtains the current value of the Hubble parameter as $H_0=74.2\pm3.6$ km s$^{-1}$ Mpc$^{-1}$, while the systematic uncertainties have been greatly reduced by the unprecedented homogeneity in the periods and metallicity of these Cepheids \cite{Riess:2009pu}. Also, by using the so called "Peak Method" that takes the BAO scale as a standard ruler in the radial direction, one can obtaining three additional data at $z=0.24,0.34,\text{and }0.43$, which are model and scale independent \cite{Gaztanaga:2008xz}. The total 15 Hubble parameter data are list in Table~\ref{table:ohd}, including the data from \cite{Stern:2009ep,Riess:2009pu,Gaztanaga:2008xz,Simon:2004tf}.
\begin{table}[h]
\centering
  \begin{tabular}{c|c|c|c|c|c|c|c|c|c|c|c|c|c|c|c}
  \hline
  \hline
  z
  & $0$ & $0.1$ &$0.17$ & $0.24$ &$0.27$ &$0.34$ &$0.4$ &$0.43$ &$0.48$ &$0.88$ &$0.9$ &$1.30$ &$1.43$ &$1.53$ &$1.75$   \\
  \hline
  \hline
  $H(z)_{obs}$ (km s$^{-1}$ Mpc$^{-1}$)
  & $74.2$ &$69$ & $83$ & $79.69$ &$77$ & $83.8$&$95$ & 86.45&$97$ &$90$&$117$ &$168$ &$177$ &$140$ &$202$     \\
  \hline
  $1\sigma$ uncertainty
  & $\pm3.6$ & $\pm12$ &  $\pm8$ &$\pm2.32$ &$\pm14$ &$\pm2.96$ &$\pm17$ &$\pm3.27$ &$\pm60$ &$\pm40$ &$\pm23$ &$\pm17$ & $\pm18$ &$\pm14$ &$\pm40$         \\
  \hline
  \end{tabular}
  \caption{\label{table:ohd} The observational Hubble parameter data. }
\end{table}

The $\chi^2 $ function of the Hubble parameter is then given by
\begin{equation}
  \chi_H^2 =  \sum^{15}_{i=1} \frac{\left[H(z_i)_{obs} - H(z_i)_{th} \right]^2}{\sigma^2(z_i)} \,.
\end{equation}

\section{Results}\label{sec:result}

\begin{figure}[h]
\begin{center}
\includegraphics[width=0.8\textwidth,angle=0]{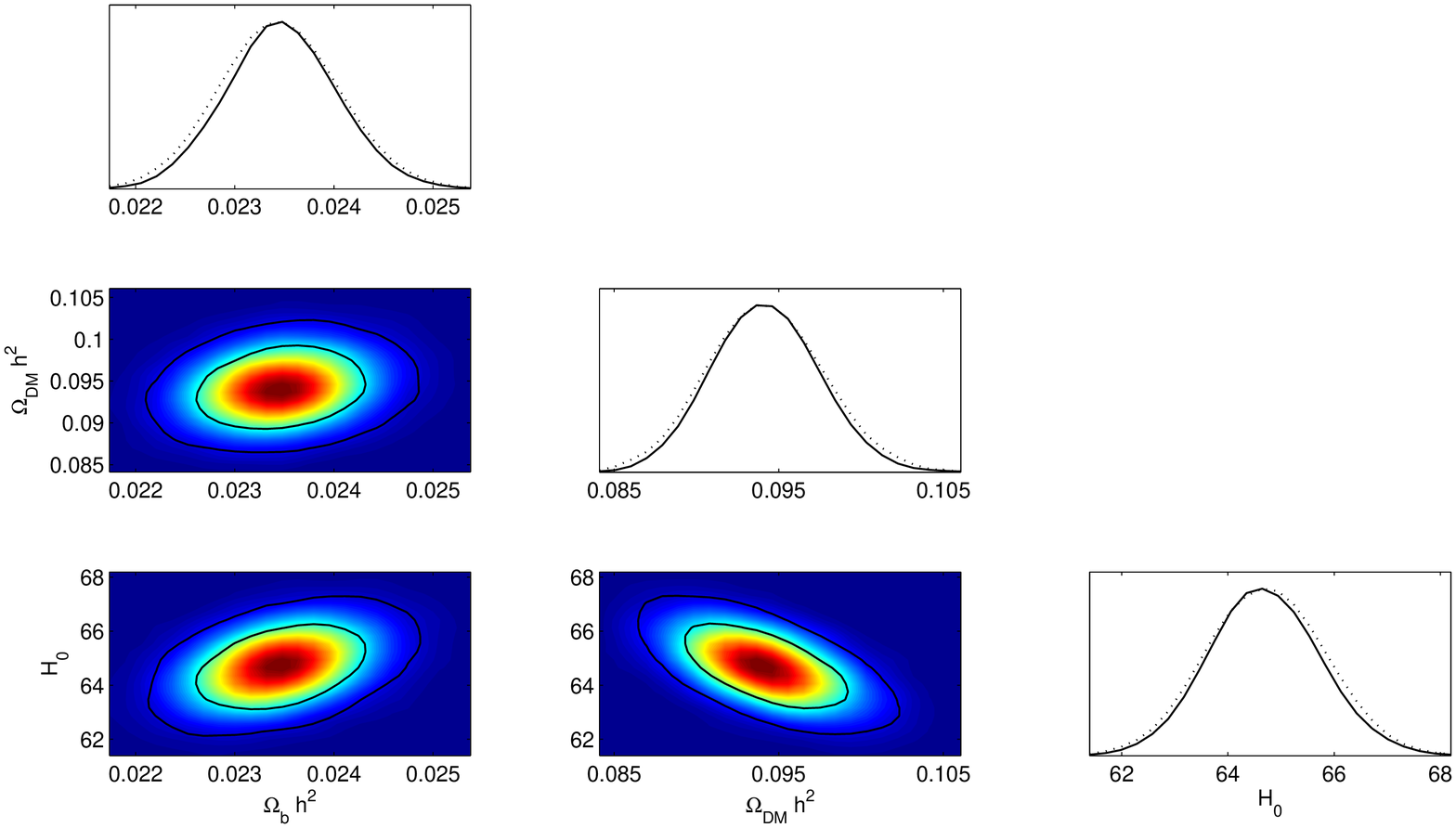}
\caption{\label{fig::ntau21} The $1$-D constraints on each parameters $(\Omega_{b0}h^2, \Omega_{m0}h^2, H_0)$, and $2$-D contours on these parameters with $1\sigma, 2\sigma$ confidence level each other in the GDE model. The data we used are Union2.1 + CMB + BAO + OHD with BBN constraint. Here the dotted curves in each $1$-D plots are the mean likelihood of the samples, while the solid curves are the marginalized probabilities for each parameters. }
\end{center}
\end{figure}

\begin{figure}[h]
\begin{center}
\includegraphics[width=0.8\textwidth,angle=0]{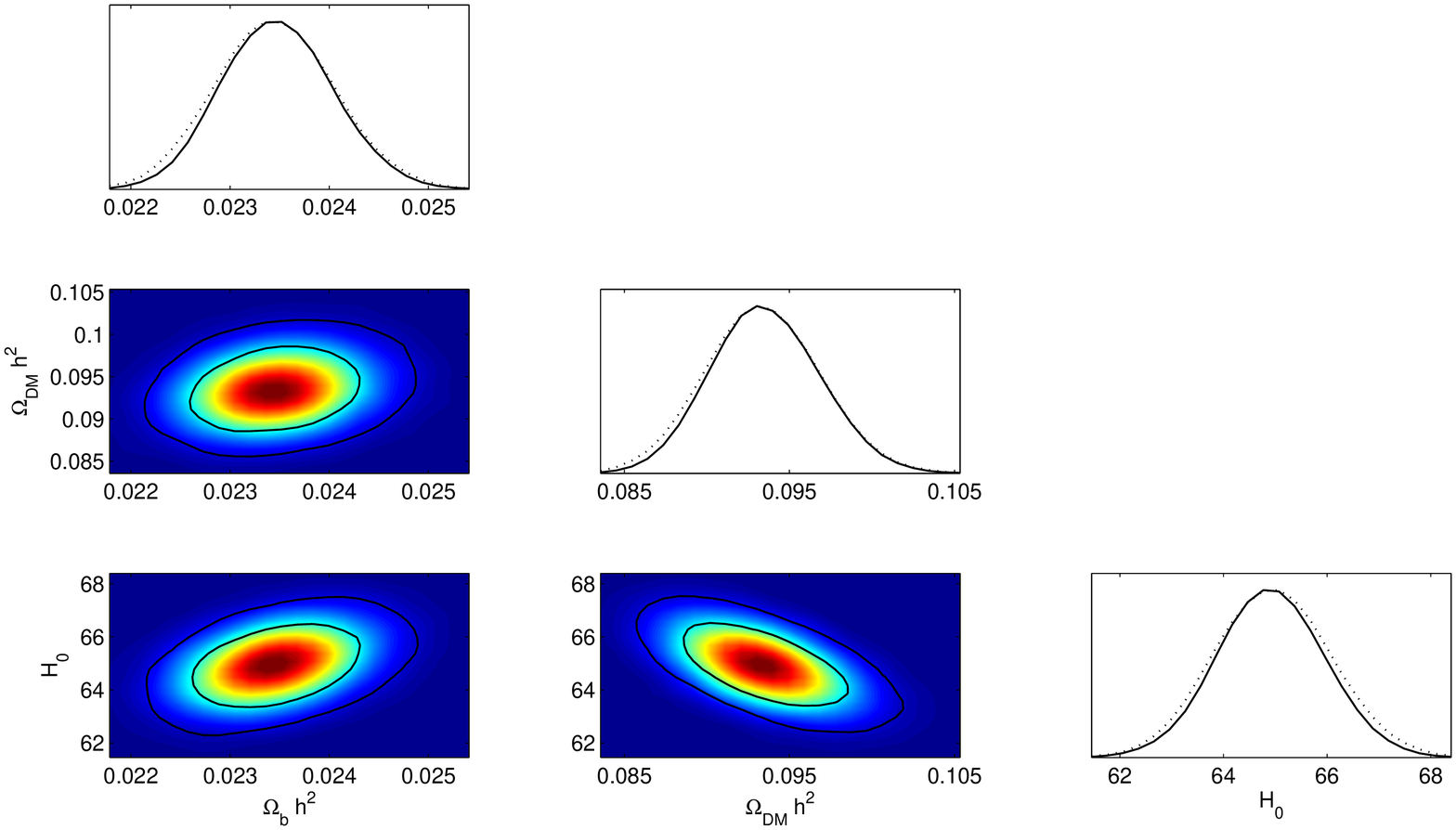}
\caption{\label{fig::ntau20}The $1$-D constraints on each parameters $(\Omega_{b0}h^2, \Omega_{m0}h^2, H_0)$, and $2$-D contours on these parameters with $1\sigma, 2\sigma$ confidence level each other in the GDE model. The data we used are Union2 + CMB + BAO + OHD with BBN constraint. Here the dotted curves in each $1$-D plots are the mean likelihood of the samples, while the solid curves are the marginalized probabilities for each parameters.}
\end{center}
\end{figure}

\begin{figure}[h]
\begin{center}
\includegraphics[width=0.8\textwidth,angle=0]{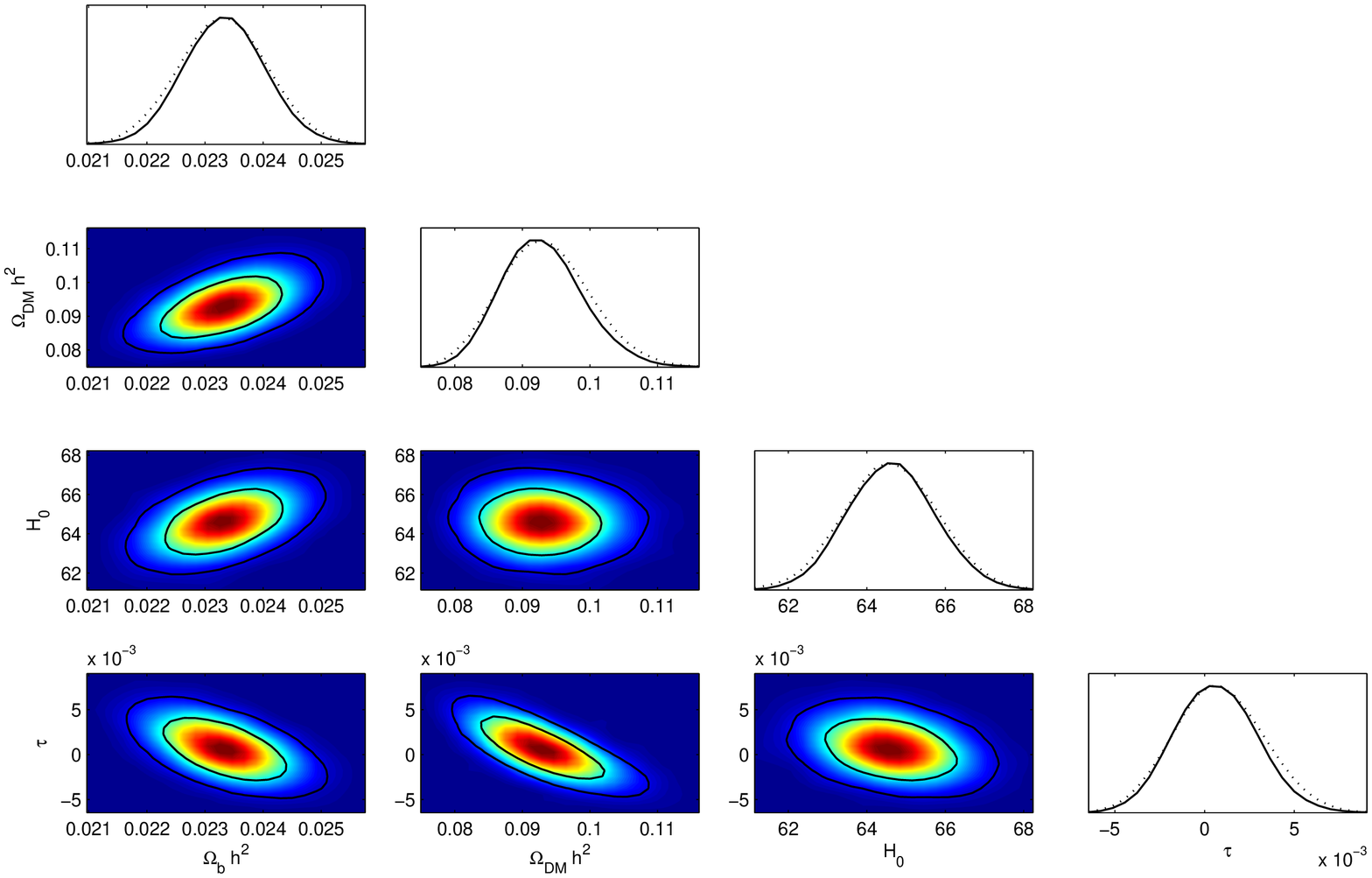}
\caption{\label{fig::tau21} The $1$-D constraints on each parameters $(\Omega_{b0}h^2, \Omega_{m0}h^2, H_0)$, and $2$-D contours on these parameters with $1\sigma, 2\sigma$ confidence level each other in the VGDE model. The data we used are Union2.1 + CMB + BAO + OHD with BBN constraint. Here the dotted curves in each $1$-D plots are the mean likelihood of the samples, while the solid curves are the marginalized probabilities for each parameters.}
\end{center}
\end{figure}

\begin{figure}[h]
\begin{center}
\includegraphics[width=0.8\textwidth,angle=0]{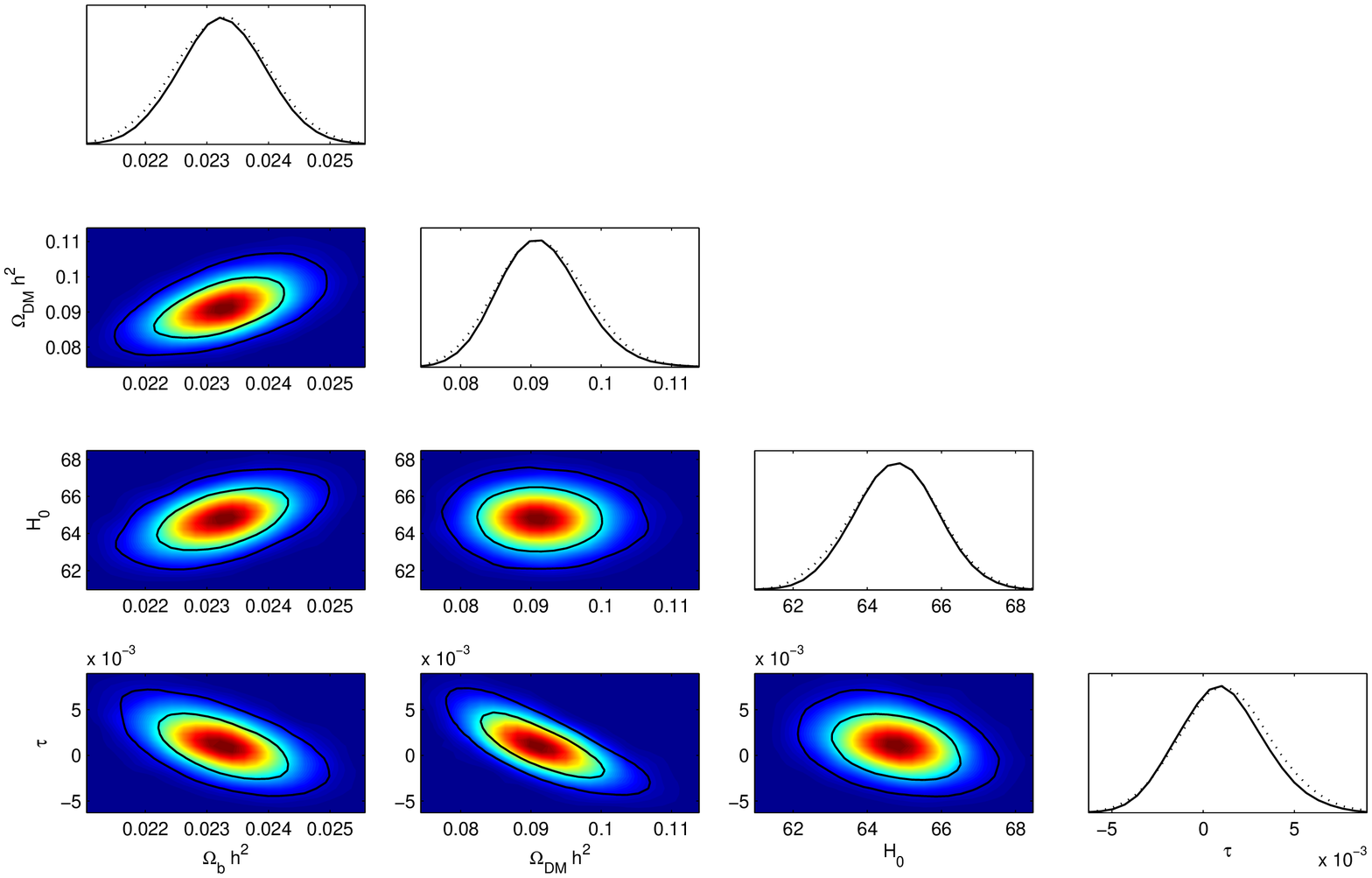}
\caption{\label{fig::tau20} The $1$-D constraints on each parameters $(\Omega_{b0}h^2, \Omega_{m0}h^2, H_0)$, and $2$-D contours on these parameters with $1\sigma, 2\sigma$ confidence level each other in the VGDE model. The data we used are Union2 + CMB + BAO + OHD with BBN constraint. Here the dotted curves in each $1$-D plots are the mean likelihood of the samples, while the solid curves are the marginalized probabilities for each parameters.}
\end{center}
\end{figure}

We now apply the maximum likelihood method for the dark energy model and the best fit parameter values with $1\sigma, 2\sigma$ errors and the corresponding values of $\chi^2_{\min}$  will be summarized in Table~\ref{table:best}. In the Fig.~\ref{fig::ntau21} and Fig.\ref{fig::ntau20}, we show the one dimensional probability distribution of each parameters in the GDE model by using the Union2.1 and Union2 supernova data with CMB, BAO, OHD data respectively, also the two dimensional contour plots between each other are shown in both figures. And the corresponding plots for VGDE model are shown in Fig.~\ref{fig::tau21} and Fig.~\ref{fig::tau20}. Comparing the GDE model with the VGDE model under the same observational data, the difference of $\chi^2_{min}$ is not so obvious and the constraint on the viscous parameter $\tau$ is very loose.
\begin{table}[h]
\centering
  \begin{tabular}{c|c|c|c|c}
  \hline
  \hline
  \multicolumn{1}{c|}{Model \& Parameter} & \multicolumn{2}{c|}{GDE} & \multicolumn{2}{c}{VGDE} \\
  \hline
  \multirow{2}{*}{Data} & Union2.1 & Union2 &Union2.1 &Union2                             \\
  \cline{2-5}
  & \multicolumn{4}{c}{ + CMB+ BAO + OHD}\\
  \hline
  \hline
  $\Omega_{b0}$ & $0.0234^{+0.0020+0.0030}_{-0.0010-0.0024}$  & $0.0234^{+0.0020+0.0029}_{-0.0009-0.0024}$
                & $0.0233^{+0.0029+0.0041}_{-0.0014-0.0035}$  & $0.0233^{+0.0029+0.0042}_{-0.0015-0.0035}$                            \\
  \hline
  $\Omega_{m0}$ & $0.0941^{+0.0117+0.0173}_{-0.0058-0.0142}$   & $0.0932^{+0.0115+0.0173}_{-0.0055-0.0141}$
                & $0.0926^{+0.0248+0.0346}_{-0.0111-0.0285}$   & $0.0910^{+0.0241+0.0339}_{-0.0109-0.0274}$                      \\
  \hline
  $H_0$         & $64.69^{+3.791+5.542}_{-1.882-4.656}$        & $64.88^{+3.739+5.622}_{-1.804-4.667}$
                & $64.51^{+4.530+6.489}_{-2.133-5.486}$        & $64.89^{+4.567+6.489}_{-2.402-5.471}$                       \\
  \hline
  $\tau$ & - & -
                & $0.0007^{+0.0095+0.0135}_{-0.0046-0.0112}$   &   $0.0010^{+0.0097+0.0137}_{-0.0045-0.0114}$                        \\
  \hline
  \hline
  $\chi^2_{min}$&588.882  & 578.252 & 588.816 & 578.028                         \\
  \hline
  \hline
  \end{tabular}
  \caption{\label{table:best} The fitting results of the parameters with $1\sigma, 2\sigma$ regions in GDE and VGDE models. Here the Hubble parameter $H_0$ is in the unit of km s$^{-1}$ Mpc$^{-1}$. }
\end{table}

\section{Discussion and Conclusion}
In summary, in this paper we have obtained the evolution of the Universe in both of the models and got the future behavior of the viscous ghost dark energy  model by using the dynamical analysis. We also used the SNeIa, CMB, BAO and OHD data with BBN constraint to constrain parameters of the ghost dark energy model with and without viscosity by using the Markov Chain Monte Carlo (MCMC) approach. The fitting results are presented in Table \ref{table:best} and Fig.~\ref{fig::ntau21}-\ref{fig::tau20}, in which we have shown the best fit parameter values with $1\sigma, 2\sigma$ errors and the corresponding values of $\chi^2_{\min}$. From the results one can see that the difference of $\chi^2_{min}$ is not so obvious for the GDE model and the VGDE model under the same observational data and the constraint on the viscous parameter $\tau$ is very loose. So, the latest observational data can not distinguish these models at this classical level. In other words, they predict almost the same evolution history of the universe and we need to take the perturbation of universe into account that will be studied in our further work.

In fact, the minimal of $\chi^2$ in eq.(\ref{tot chi2}) is very sensitive to the observational error of the distance modulus. Once the error is reduced in the future, then one may distinguish these models and even rule out some of them. Thus, more precise data are very needed. We also hope that future observation data could give more stringent constraints on the parameters in the ghost dark energy model, especially on the viscous parameters.

\acknowledgments
This work is supported by National Science Foundation of China grant Nos.~11105091 and~11047138, National Education Foundation of China grant  No.~2009312711004, Shanghai Natural Science Foundation, China grant No.~10ZR1422000,  and  Shanghai Special Education Foundation, No.~ssd10004.
The data fitting is based on the public available \textbf{CosmoMC} package.
\appendix

\end{document}